\documentclass[pra,aps,twocolumn,groupedaddress,letterpaper,showpacs,amssymb,amsmath,10pt]{revtex4-1}
\usepackage{graphicx}
\usepackage[dvips]{epsfig}
\usepackage{dcolumn}
\usepackage{bm}
\usepackage{array}
\usepackage{amssymb}

\usepackage{natbib}

\newcommand{\bi}{\begin{itemize}}
\newcommand{\ei}{\end{itemize}}
\newcommand{\be}{\begin{equation}}
\newcommand{\ee}{\end{equation}}
\newcommand{\ba}{\begin{eqnarray}}
\newcommand{\ea}{\end{eqnarray}}

\newcommand{\upa}{\uparrow}
\newcommand{\dna}{\downarrow}

\newcommand{\COMMENTED}[1]{}

\newcommand{\ket}[1]{|#1\rangle}

\newcommand{\bfr}{\mathbf{r}}

\begin{document}

\title{Itinerant ferromagnetism in a Fermi gas with contact interaction:\\
Magnetic properties in a dilute Hubbard model}

\author{Chia-Chen Chang${}^1$}

\author{Shiwei Zhang${}^{1,2,3}$}

\author{David M. Ceperley${}^4$}

\affiliation{
${}^1$Department of Physics, College of William and Mary, Williamsburg, VA 23187, USA\\
${}^2$Institute of Physics, Chinese Academy of Sciences, Beijing 100190, China\\
${}^3$Department of Physics, Renmin University of China, Beijing 100872, China\\
${}^4$Department of Physics, University of Illinois at Urbana-Champaign, Urbana, IL 61801, USA
}

\begin{abstract}

Ground state properties of the repulsive Hubbard model on a
cubic lattice are investigated by means of the auxiliary-field
quantum Monte Carlo method. We focus on low-density systems
with varying on-site interaction $U/t$, as a model relevant to
recent experiments on itinerant ferromagnetism in a dilute
Fermi gas with contact interaction. Twist-average boundary
conditions are used to eliminate open-shell effects and large
lattice sizes are studied to reduce finite-size effects. The
sign problem is controlled by a generalized constrained path
approximation. We find no ferromagnetic phase transition in
this model. The ground-state correlations are  consistent with
those of a paramagnetic Fermi liquid.

\end{abstract}

\pacs{71.10.Fd,02.70.Ss}

\maketitle

{\it Introduction} -
The study of ferromagnetism has a long history in physics.
At the microscopic level, the formation of ferromagnetic order
is a consequence of strong interactions.
Heisenberg first pointed out that
an exchange interaction that lowers the energy of a pair
of parallel spins would favor ferromagnetism.
However a localized-spin mechanism cannot be fully
responsible for ferromagnetism in transition
metals, for instance iron and nickel, where electrons are extended.
Both the interactions and the delocalized nature of electrons
have to be taken into account at a more fundamental level.

A generic description of itinerant ferromagnetism is given by
the three-dimensional (3-D) Hubbard model \cite{Hubbard1963}: \be
  H=-t\sum_{\langle ij\rangle \sigma}\left(
    c_{i\sigma}^\dagger c_{j\sigma} + h.c. \right)
    + U\sum_i n_{i\upa} n_{i\dna}.    \label{eq:Hubbard}
\ee The operator $c_{i\sigma}^\dagger$ ($c_{i\sigma}$) creates
(annihilates) an electron with spin $\sigma$
($\sigma=\upa,\dna$ ), $i$ enumerates the sites in an
$N=L^3$ lattice, and $\langle ij \rangle$ denotes a sum of nearest
neighbor pairs. The parameter $t$ is the nearest-neighbor
hopping amplitude, and $U>0$ is the on-site interaction
strength. The total density is $n=(N_\uparrow+N_\downarrow)/N$.
This simple Hamiltonian
contains both the itinerant character 
and local
repulsion. However, because neither of the two terms alone
favors ferromagnetic ordering, the magnetic correlations in the
Hubbard model is not obvious.

The first evidence of ferromagnetism in the Hubbard model was discussed
by Nagaoka \cite{Nagaoka1966} and by Thouless \cite{Thouless1965}.
They showed that the ground state with a single hole
in any finite
bipartite lattice with $U\rightarrow \infty$ (and periodic
boundary conditions) is fully polarized.
Subsequent studies indicate that
the stability of the state
with more holes can depend on system size
\cite{Takahashi1982,*Doucot1989,*Zhang1991,*Puttika1992},
and boundary conditions \cite{Riera1989,*Barbieri1990}.
The critical doping for the onset of
ferromagnetism is still
an open question \cite{Puttika1992,Becca2001,Carleo2010}.
Away from infinite-$U$, the existence of ferromagnetism at
non-zero density is less certain. Whether ferromagnetism is a
generic property of the Hubbard model is still not answered.

Rapid experimental progress in cold atoms has opened a new
avenue for exploring the physics of itinerant ferromagnetism.
In a recent experiment aimed to simulate the Stoner Hamiltonian
(i.e. spin 1/2 fermions in continuous space interacting with a
repulsive contact potential), a dilute gas of two hyperfine
states of ${}^6$Li atoms are tuned to interact via large
positive scattering lengths. Signatures of ferromagnetic
instability \cite{Jo2009} have generated a lot of theoretical
interest
\cite{Zhai2009,*Cui2010,Pilati2010,Chang2010,Pekker2010}.

The Hubbard Hamiltonian in Eq.~(\ref{eq:Hubbard}) gives a
reasonable representation of the Stoner Hamiltonian on a
lattice. As the density $n\rightarrow 0$ it seems clear that no
ferromagnetism exists in the model in Eq.~(\ref{eq:Hubbard}) \cite{Tasaki1998},
since the maximum scattering length is bounded by
$\sim 1/3.173$ lattice spacing \cite{Castin2004,Purwanto2005}.
However, at low but not zero
density, the magnetic properties and the
phase diagram
of the 3-D Hubbard model are not clear. We address this question
here by very accurate many-body simulations with the
constrained path Monte Carlo (CPMC) method.

{\it Method} - The CPMC method
\cite{Zhang1995,Zhang1997,Zhang2003,Chang2008} projects the
many-body ground state $\ket{\Psi_0}$ from a trial wave
function $\ket{\Psi_T}$ by repeated application of an
imaginary-time propagator $e^{-\Delta\tau H}$ ($\Delta\tau$ is
the Trotter time step), provided that $\ket{\Psi_T}$ satisfies
$\langle\Psi_0|\Psi_T\rangle\neq 0$.
The propagator is decomposed into
$e^{-\Delta\tau H}\approx e^{-\Delta\tau H_1/2}e^{-\Delta\tau H_2}e^{-\Delta\tau H_1/2}
+\mathcal{O}(\Delta\tau^3)$,
where $H_1$ and $H_2$ are one- and two-body parts of $H$ respectively.
The two-body part $e^{-\Delta\tau H_2}$ is
further decoupled into a sum over one-body projectors
in Ising fields \cite{Hirsch1985}.
This leads to a formally exact expression
$ e^{-\Delta\tau H}=\sum_{\{\mathbf{x}\}}P(\{\mathbf{x}\})B(\{\mathbf{x}\})$,
where $\{\mathbf{x}\}$ is a collection of $N$ Ising fields,
$P(\{\mathbf{x}\})$ is their probability distribution, and
$B(\{\mathbf{x}\})$ is a one-body projector.
The multidimensional summation is carried out efficiently by
importance-sampled random walks with non-orthogonal Slater determinants (SDs),
where the one-body projectors $B(\{\mathbf{x}\})$ propagate one SD into another.

The fermion sign problem is controlled approximately
by the constrained path approximation.\cite{Zhang1995,*Zhang1997}
The many-body ground
state is given by $|\Psi_0\rangle = \sum_\phi w(\phi)
|\phi\rangle$, where $|\phi\rangle$ are SDs sampled by the QMC,
and their probability distribution determines the weight
factors $w(\phi)$.
Because the Schr\"odinger equation is linear, $|\Psi_0\rangle$
is degenerate with $-|\Psi_0\rangle$. 
In a random walk, the SDs can
move back and forth between the two sets of solutions. The
appearance of the two sets with opposite signs in the Monte
Carlo samples is the origin of the sign problem.
To control the problem, the walker is required to satisfy the
constraint $\langle\Psi_T|\phi\rangle >0$ in the course of the
random walk. This is the only approximation in our method. More
formal discussions of the theoretical basis of the generalized
constrained path approximation and benchmarks can be found
elsewhere \cite{Zhang1995,*Zhang1997,Zhang2003}.
In the Hubbard model, the energy at $U=4t$ is typically within
$< 0.5\%$ of the exact diagonalization result \cite{Chang2008}.
Extensive benchmarks of this approach for molecules and solids
are in Refs. \cite{AlSaidi2006,*AlSaidi2007,Purwanto2009}.

The constrained path approximation is similar in spirit to the
fixed-node approximation in the diffusion Monte Carlo (DMC)
method \cite{Ceperley1980,Foulkes2001}, which has been used for
all recent simulation work on the problem of itinerant
ferromagnetism in the Stoner model
\cite{Conduit2009a,Pilati2010,Chang2010}.
In fixed-node DMC one uses a real-space trial function
$\Psi_T(R)$ to determine the sign
of the ground-state
wave function. The random walks, which involve movements of electron
coordinates $R$ (a $3\,(N_\uparrow+N_\downarrow)$-dimensional vector),
are constrained to the region where $\Psi_T(R)>0$.
Since, in CPMC the random walks take place in the space of SDs,
where fermionic statistics are automatically maintained, the
sign problem is reduced.
As a result, the constrained path approximation is less
sensitive to $|\Psi_T\rangle$ and typically has smaller
systematic errors.

In this work we apply twist-averaged boundary conditions
(TABCs) \cite{Lin2001}. Under TABCs, the wave function gains a
phase when electrons wind around the periodic boundary
conditions: $
  \Psi(\ldots,\mathbf{r}_j+\mathbf{L},\ldots) = e^{i\widehat{\mathbf{L}}\cdot{\bm\Theta}}
  \Psi(\ldots,\mathbf{r}_j,\ldots),
$
where $\widehat{\mathbf{L}}$ is the unit vector along $\mathbf{L}$,
and ${\bm\Theta}=(\theta_x,\theta_y,\theta_z)$ are random twists
over which we average. A simple generalization of the CPMC method 
can be made to handle
the overall phase that arises from TABC \cite{Zhang2003,Chang2008}.
As an additional benchmark for the present work, we studied
several low-density $L=4$ systems in detail. For example, at
$U=16t$ with $n=0.25$, the CPMC energy, averaged
over $1000$ ${\bm\Theta}$-points, agrees to better than $0.2$\%
with exact diagonalization.


\begin{center}
\begin{figure}
\includegraphics[scale=0.285]{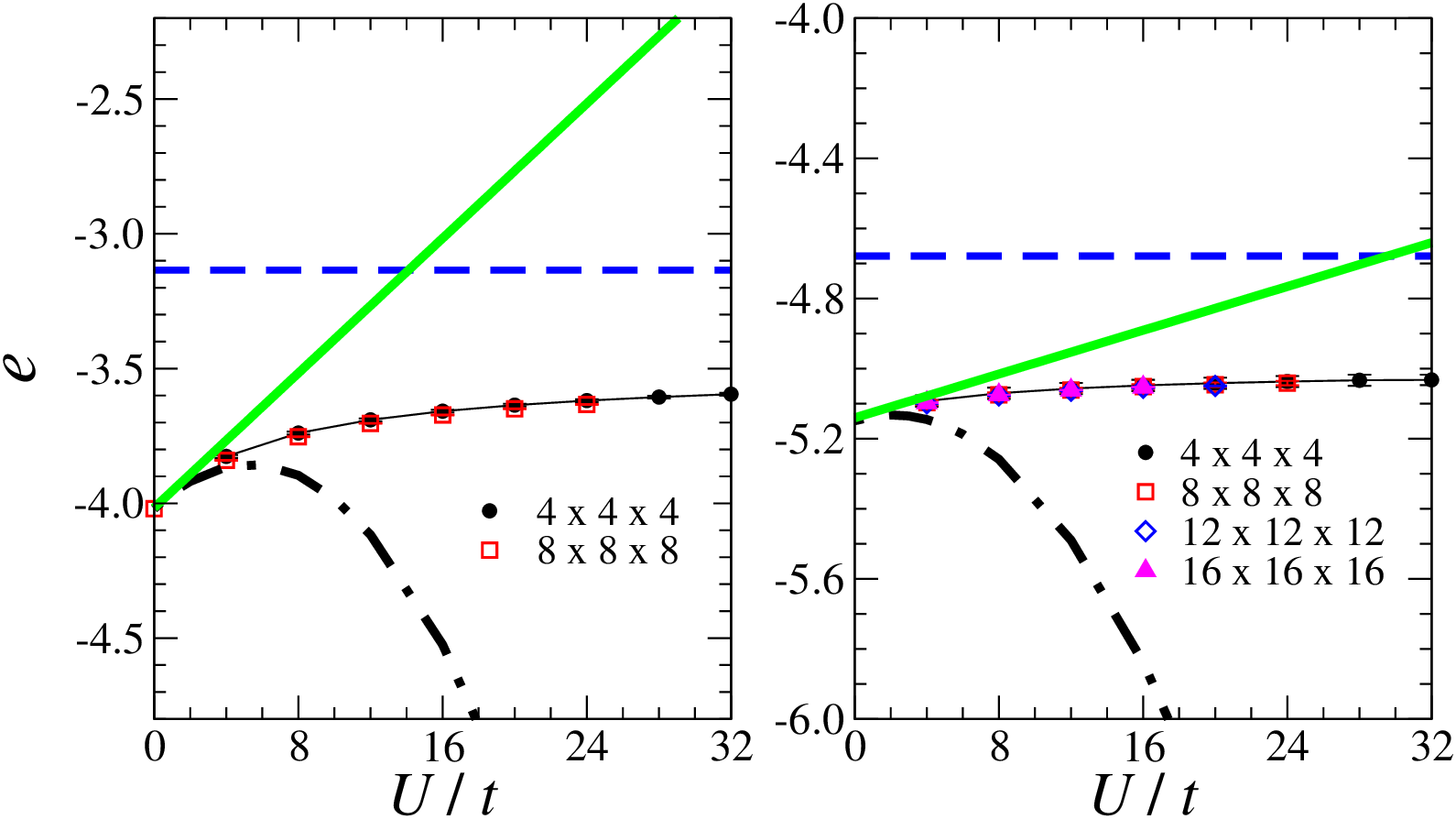}
\caption{(color online). Ground state energy per particle $e$
as a function of interaction strength $U/t$ at $n=0.25$ (left)
and $n=0.0625$ (right). Symbols represent $e_{CPMC}$. Dashed
(blue) line corresponds to the energy of a saturated
ferromagnetic state ($e_{FM}$).
$e_{MF}$ energy is represented by the thick solid (green) line.
$e_P$ (perturbation theory \cite{Metzner1989}) is plotted
by dot-dashed line.
\label{fig:ener-gr}
}
\end{figure}
\end{center}

{\it Energy} - We first compare the ground-state energy of an
unpolarized system ($N_\uparrow=N_\downarrow$) with that of a
fully polarized state at the same total density. The results
are summarized in Fig.~\ref{fig:ener-gr}.
Because electrons of the same spin do not interact,
the energy of the fully polarized state, $e_{FM}$, is purely kinetic and does not depend on $U$.
In mean-field (MF) theory, the energy of a system with $n_\sigma=N_\sigma/N$
(with $n=n_\uparrow+n_\downarrow$) is
$
  e_{MF}(U,n) = [e_0(n_\uparrow) n_\uparrow+e_0(n_\downarrow) n_\downarrow 
                       + Un_\uparrow n_\downarrow]/n,
$ 
where $e_0(n_\sigma)$ is the energy of the fully polarized system 
at density $n_\sigma$.
At $n=0.25$, MF predicts a paramagnetic to ferromagnetic phase
transition at $U = 13.9t$. 
This is to be compared to the corresponding transition point
$k_F a\sim \pi/2$ in the continuum Stoner Hamiltonian, where
$k_F = (3\pi^2 n)^{1/3}$ is the Fermi wave vector, and $a$ is
the scattering length in continuum. 
When the system density is lowered to $n=0.0625$, the MF
transition in the Hubbard model is at a larger interaction,
$U = 29.3 t$. 
The equation of state has also been obtained from perturbation
theory for an unpolarized system \cite{Metzner1989}: $
  e_{P}(U,n) = e_{MF}(U,n) + e_c(U,n).
$ The last term, $e_c(U,n)$, is the correlation energy
estimated to $\mathcal{O}(U^2)$. 
The result is also included in
Fig.~\ref{fig:ener-gr}.

The CPMC result for the ground state energy $e_{CPMC}$ is
obtained by averaging over twist-angles. The
energies calculated from different lattice sizes are shown by
different symbols in Fig.~\ref{fig:ener-gr}. It can be seen
that our remaining finite-size errors are negligible on this
scale. Free-electron trial wave functions are used for the
constraint. In a few cases we have also checked with
unrestricted Hartree-Fock trial wave functions, which gave
statistically indistinguishable CPMC energies. The energies
shown are for finite time steps, with $\Delta\tau$ satisfying
$U\Delta\tau< 0.2$. The residual Trotter error is
${\cal{O}}(10^{-2})$, smaller than the symbol size.

We see that MF theory, which gives a reasonable estimate of the
energy at small $U$, quickly shows severe deviations as the
interaction becomes stronger. The perturbation result,
$e_{P}(U,n)$, gives an improved estimate of energy for small
$U$, but deviates once the system enters the
intermediate interaction regime $U \gtrsim 5t$. At the MF
transition point, the CPMC energy is significantly lower than
$e_{FM}$. Indeed the CPMC energy remains lower than $e_{FM}$
across the entire range of $U$ simulated. No indication of a
ferromagnetic transition is seen.

\begin{center}
\begin{figure}
\includegraphics[scale=0.29]{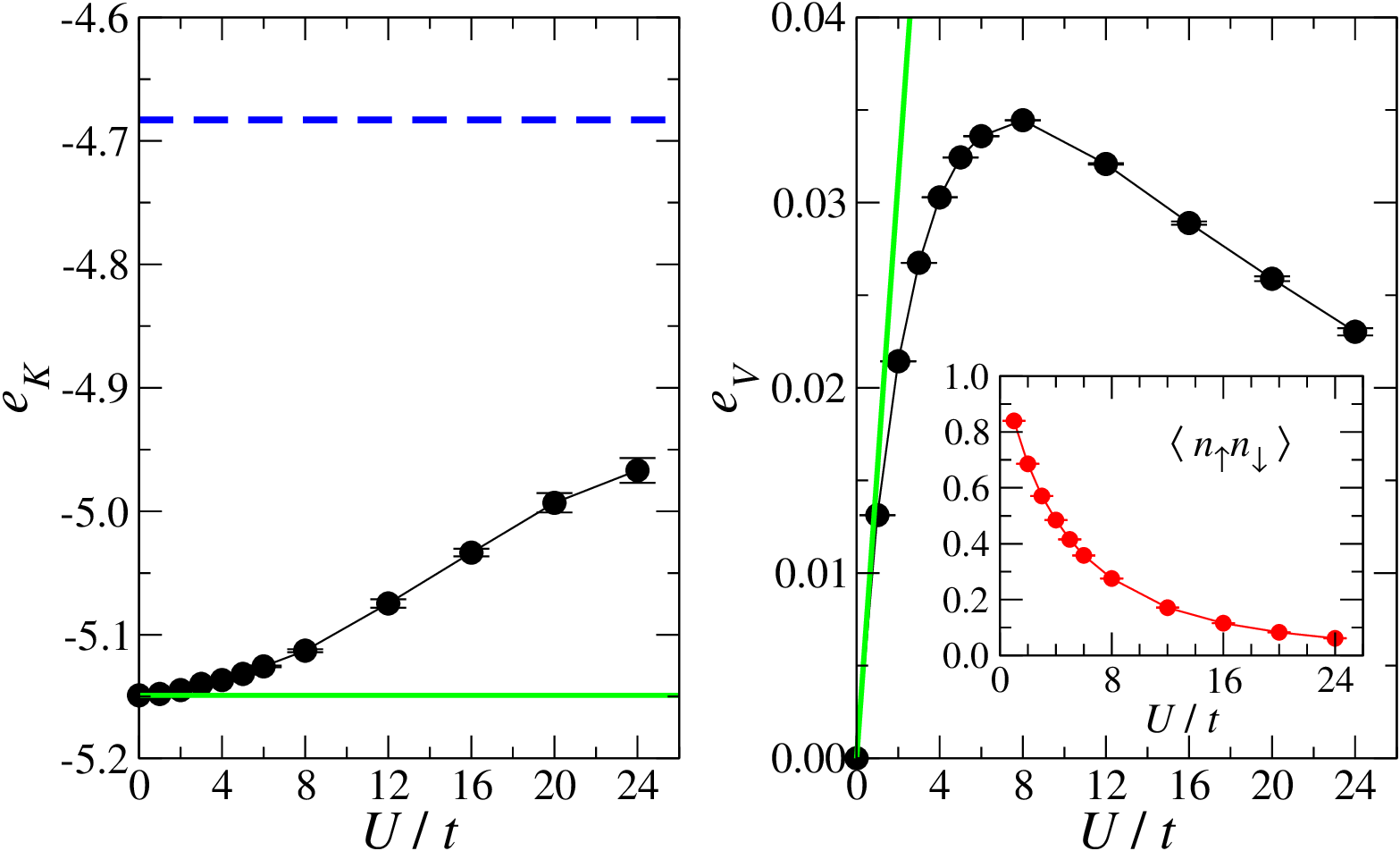}
\caption{(color online). Kinetic (left panel) and interaction
(right panel) energies as a function of interaction strength at
density $n=0.0625$. Symbols are the CPMC data obtained on an
$8^3$ lattice. Lines are defined in the same way as in
Fig.~\ref{fig:ener-gr}. The inset on the right shows the double
occupancy, normalized to $1$ at $U=0$. } \label{fig:ener-ek-ev}
\end{figure}
\end{center}

Individual components of the energy are shown in
Fig.~\ref{fig:ener-ek-ev}. As $U$ increases, electrons in the
unpolarized system occupy higher momentum states, outside the
Fermi level, which increases the kinetic energy compared to the
MF result. This enables the system to drastically decrease the
interaction energy, by suppressing double occupancy. The net
effect is that the total energy is greatly reduced and remains
below $e_{MF}$ and $e_{FM}$.

{\it Correlation function} -
To probe the nature of the ground state, we examine the
spin-dependent pair correlation function: \be
  g_{\sigma\sigma'}(\bfr)=\frac{1}{\bar{n}_\sigma \bar{n}_{\sigma'}}\frac{1}{N}
  \sum_{\bfr'}
  \langle
      n_{\bfr+\bfr',\sigma}\cdot n_{\bfr',\sigma'}
  \rangle. \label{eq:spin-corr}
\ee
The CPMC expectations are evaluated by the back-propagation
technique \cite{Zhang1997,Purwanto2004}. We average over
different $\mathbf{r}$'s to obtain $g_{\sigma\sigma'}(r)$, with
$r\equiv |\mathbf{r}|$ since the correlation function is primarily a function of
distance in the paramagnetic or ferromagnetic phases.

\begin{center}
\begin{figure}
\includegraphics[scale=0.31]{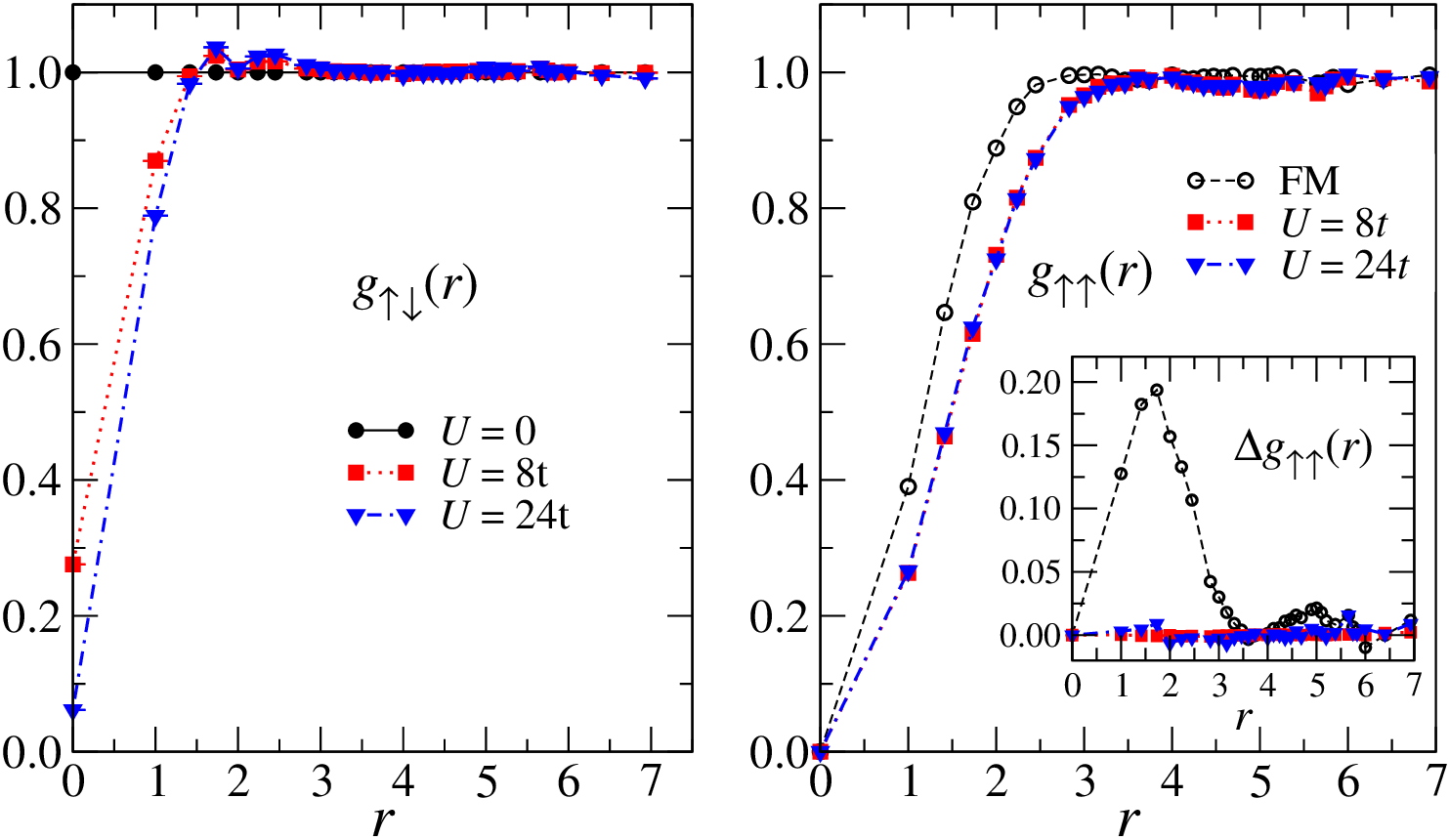}
\caption{(color online). 
Left: Anti-parallel spin-spin pair correlation function of
an unpolarized state at different interaction strengths.
Right: Comparison of the parallel pair correlation
functions of the fully polarized state (FM) and the unpolarized
states at different interaction strengths. 
The inset shows
$\Delta g_{\uparrow\uparrow}(r)=g_{\uparrow\uparrow}(r)-g^0_{\uparrow\uparrow}(r)$,
where $g^0_{\uparrow\uparrow}(r)$ is the correlation function of
the unpolarized non-interacting system.
In both panels, the system is an $8^3$ lattice at density $n=0.0625$.
}
\label{fig:corr}
\end{figure}
\end{center}

The anti-parallel pair correlation
$g_{\uparrow\downarrow}(r)$ is a constant in a non-interacting system
or in the MF solution.
In the presence of interaction,
a correlation hole is created surrounding each electron.
At $n=0.0625$, the size of the correlation hole is $r_{cor}\lesssim\sqrt{3}$.
As $U$ is increased, the correlation hole
becomes deeper, 
as illustrated in the left panel in Fig.~\ref{fig:corr}.
Compared to $g_{\uparrow\downarrow}(r)$, the change in
the parallel-spin pair correlation 
$g_{\uparrow\uparrow}(r)$ is less dramatic from the MF or non-interacting
result.
Strong interaction does appear to increase $g_{\uparrow\uparrow}(r)$
slightly at short distance. However, the correlation remains much less than
that in the FM case.

\begin{center}
\begin{figure}
\includegraphics[scale=0.3]{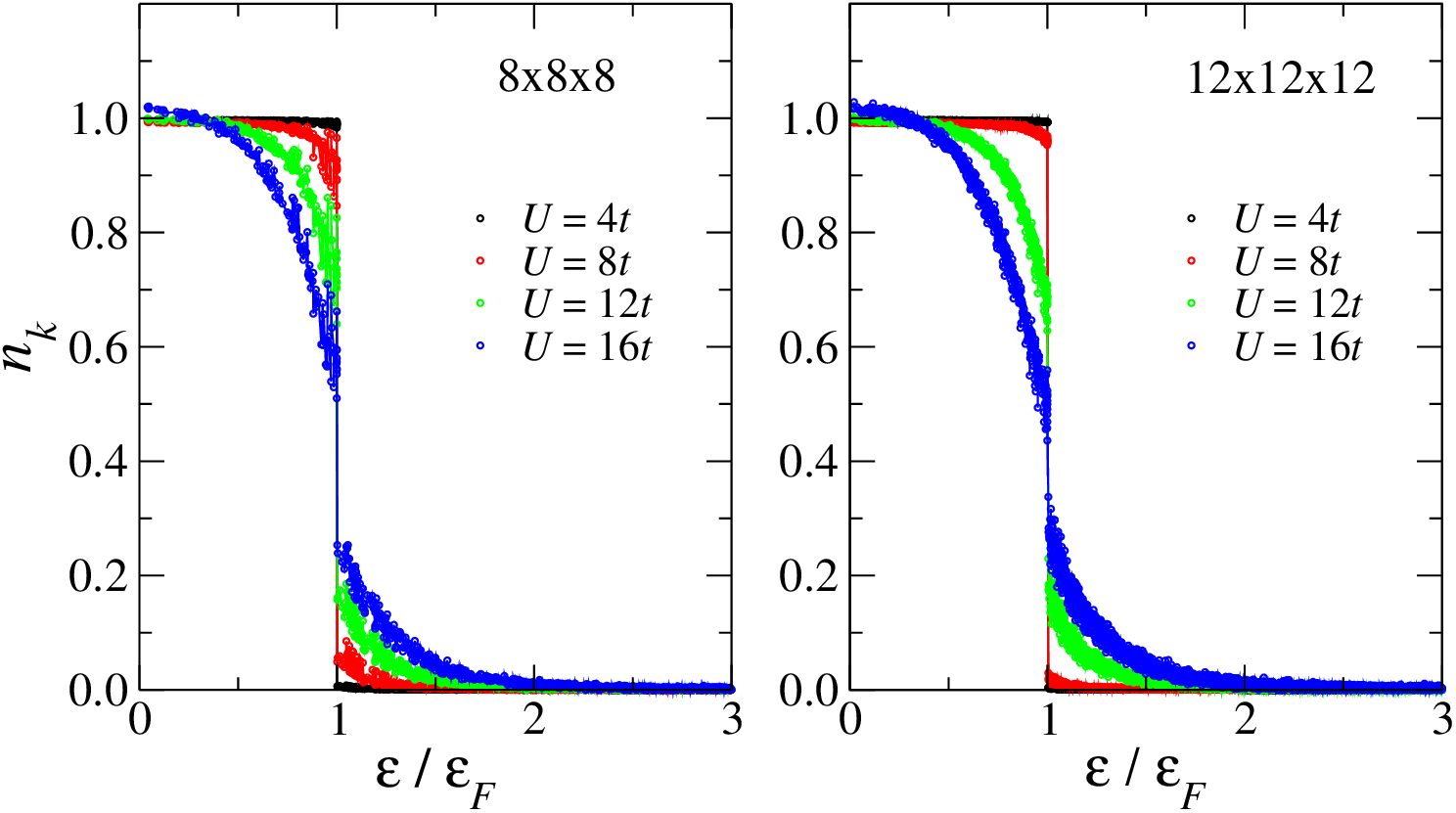}
\caption{The momentum distribution $n_\mathbf{k}$ for values
of $U$ plotted as a function of the
single particle energy. Two lattice sizes are shown, at the
density $n=0.0625$. In each system, we average over 10 random 
twist angles.
}
\label{fig:nk}
\end{figure}
\end{center}

{\it Momentum distribution} -
The creation of correlation hole is a result
of minimizing the interaction
energy. Electrons of opposite spins rearrange their relative positions
to reduce the potential energy.
The cost of the rearrangement is the kinetic energy increase, as
discussed earlier.
This can also be observed in Fig.~\ref{fig:nk} where the momentum distribution
$n_\mathbf{k}$ is shown for different interaction strengths.
We have plotted $n_\mathbf{k}$ as a function of the single-particle
energy level
$\varepsilon(\mathbf{k})=2\sum_{\alpha=x,y,z}
[1-\cos (k_\alpha+\Theta_\alpha/L)]$,
in units of the Fermi energy $\varepsilon_F$.
Each curve contains the result of $n_\mathbf{k}$ from multiple
$\Theta$-points.
At $U=4t$, the distribution is very close to the
non-interacting momentum distribution
with only a few low lying excitations near the Fermi surface
(FS). As $U$ is increased, more higher $\mathbf{k}$ states are
populated outside the FS.
In $n_\mathbf{k}$ a jump appears at $\varepsilon_F$ which can
be read off directly in our finite size simulations. The jump
indicates that the system is a normal Fermi liquid, with
the value of the jump proportional to the renormalization
factor $Z$. Its precise value can be determined with more
extensive calculations and finite size scaling.


{\it Discussion} -
Although we have focused on the
ground state of a homogeneous Fermi gas, it is not difficult to
extend the results to the case with an external trap.
For example, the kinetic energy results
in Fig.~\ref{fig:ener-ek-ev} indicate that, with a trap, there
would be a {\it minimum} in the curve of $e_K$ versus
interaction strength, as observed in the experiment
\cite{Jo2009} (see also discussion in Ref.~\cite{Zhai2009}).
Effects of confinement on the kinetic energy have been
investigated in detail by CPMC for trapped Bose gases
\cite{Purwanto2005}. The MF kinetic energy was shown to
decrease monotonically because the gas expands in the trap as
the scattering length $a$ is increased;
on the other hand, correlation effects lead to an increase of
$e_K$,
similar to Fig.~\ref{fig:ener-ek-ev}. This
competition results in a non-monotonic curve, with a {\it
minimum} in the kinetic energy at a finite scattering length.

The Hubbard model, of interest in its own right, contains some
of the same features (namely itinerant electrons and local
interaction) as the continuum Stoner Hamiltonian.
However, there are differences with respect to the experiment
worth emphasizing.
The experiment is in the continuum, using an attractive
(negative) interaction with an effective positive scattering
length for the excited state that describes the prepared state.
(However, there are questions whether such an effective
description is appropriate \cite{Pekker2010}.) 
In our simulation, we use a discretized
representation, with positive on-site interaction. As mentioned
before, the lattice model leads to a scattering length bounded
by roughly the lattice spacing. 
Using the above values for the maximum scattering length
and $k_F$ in the unpolarized phase, we find that
$k_F a < 1.03 n^{1/3}$. (For the transition to the
ferromagnetic phase, $n \le 1$ since one cannot have two
like-spin fermions on a single site. To focus on the dilute limit,
we have done calculations for up to $n=0.5$.)

Recently the problem of itinerant ferromagnetism in repulsive
Fermi gases has been studied by several groups
\cite{Conduit2009a,Pilati2010,Chang2010} using the 
DMC method with the fixed-node approximation.
These calculations all found the existence of a ferromagnetic
instability.
The DMC calculations were all done in the continuum, while the
present calculation is for the Hubbard (lattice) model. In the
DMC simulations, the atomic interaction is modeled by a {\em
repulsive\/} potential whose range is determined by the
scattering length.
We note that since the scattering length diverges near
resonance, the range of potential (or the range of the node in
the Jastrow when a negative interaction is used) can become
very large. Note that the hard-sphere-like interaction is only
between unlike spins. As the scattering length approaches the
interparticle spacing, there is a strong tendency to separate
into a spin-up and spin-down domains, to lower the interaction
energy; i.e.~it favors ferromagnetism.

Of lesser importance, the DMC fixed-node errors in the
calculated ground state energies bias the result in favor of
ferromagnetism, since nodal surfaces for the ferromagnetic state
are more accurate than the spin unpolarized state \cite{Zong2003}.
Although the constrained path error from our
calculations could also be biased, previous calculations
indicate \cite{Zhang1997,Zhang2003,Purwanto2009} that the
systematic error in CPMC  is smaller than the fixed-node error
from single determinant trial wave functions used in these
calculations.


{\it Summary} -
We have examined the magnetic properties in the
ground state of the dilute 3D Hubbard model, using the CPMC method and
twist-averaged boundary conditions.
Our simulation results indicate that there is
no ferromagnetic instability in this model with strong on-site
repulsions for densities up to $0.5$.
The ground state appears to be a paramagnetic Fermi liquid.
The total energy is effectively lowered by electron correlation
which, while increasing the kinetic energy, can strongly suppress double
occupancy to lower the interaction energy. In the presence of a trap,
the kinetic energy can be decreased by the expansion of the gas due to
repulsive interaction.
A kinetic energy minimum, which was observed in the experiment,
can be understood in terms of
the competitions between these effects.
We have also discussed the difference between our calculations and
recent results from DMC simulations, as well as connections and differences
with the Fermi gas itinerant ferromagnetism experiments.


{\it Acknowledgments} - C.C.~and S.Z.~acknowledge support from
ARO (56693-PH) and NSF (DMR-0535592) and D.M.C. from the OLE
program. Computations were carried out at ORNL (Jaguar XT4) and
William \& Mary (CPD and SciClone clusters). We thank Jie Xu
for help and T.L.~Ho for useful discussions.

\bibliography{./ferro}

\end{document}